\documentclass[preprint,number,sort&compress]{elsarticle}

\usepackage{booktabs}
\usepackage{siunitx}
\usepackage{amsmath}
\usepackage{amssymb}
\usepackage{graphicx}
\usepackage{xcolor}

\newcommand{\undu}{\mathrm{u}}

\begin{document}
\title{Progress on Experiments towards LWFA-driven Transverse Gradient Undulator-Based FELs}
\author[kit]{A.~Bernhard\corref{cor1}}
\ead{axel.bernhard@kit.edu}
\author[kit]{V.~Afonso Rodr\'{i}guez\fnref{fn1}}
\author[fsu,hij]{S.~Kuschel}
\author[fsu]{M.~Leier}
\author[kit]{P.~Peiffer\fnref{fn2}}
\author[fsu,hij]{A.~S\"avert}
\author[fsu,hij]{M.~Schwab}
\author[kit]{W.~Werner}
\author[kit]{C.~Widmann\fnref{fn4}}
\author[kit]{A.~Will\fnref{fn5}}
\author[kit]{A.-S.~M\"uller}
\author[fsu,hij]{M.~Kaluza}

\cortext[cor1]{Corresponding author}
\fntext[fn1]{Now at ITK Engineering GmbH, 76761 R\"ulzheim, Germany}
\fntext[fn2]{Now at Johannes-Gutenberg-Universit\"at Mainz, Germany}
\fntext[fn4]{Now at ADMEDES GmbH, Pforzheim, Germany}
\fntext[fn5]{Now at CERN, 1211 Geneva 23, Switzerland}

\address[kit]{Karlsruhe Institute of Technology, Kaiserstr.~12, 76131 Karlsruhe,
Germany}
\address[fsu]{Friedrich-Schiller-Universit\"at Jena, 07737 Jena, Germany}
\address[hij]{Helmholtz Institute Jena, Fr\"obelstieg 3, 07743 Jena, Germany}

\begin{abstract}
 Free Electron Lasers (FEL) are commonly regarded as the potential key
 application of laser wakefield accelerators (LWFA). It has been found that electron
 bunches exiting from state-of-the-art LWFAs exhibit a
 normalized 6-dimensional beam brightness comparable to those in
 conventional linear accelerators. Effectively exploiting this beneficial beam property for
 LWFA-based FELs is challenging due to the extreme initial conditions
 particularly in terms of beam divergence and energy spread. Several different
 approaches for capturing, reshaping and matching LWFA beams to suited
 undulators, such as bunch decompression or transverse-gradient undulator
 schemes, are currently being explored. In
 this article the transverse gradient undulator concept will be discussed with a focus on recent
 experimental achievements.
\end{abstract}
\maketitle

\section{Introduction}
Due to the extremely high longitudinal electric fields present in laser- or
beam-driven  plasma waves,
plasma wakefield accelerated electron bunches can gain energies sufficient to
generate synchrotron radiation in the X-ray regime within only a few millimetres
to centimeters of acceleration length. It is common sense that fully exploiting
the potential of laser-driven plasma accelerators could bring
laboratory scale, cost efficient, highly brilliant radiation sources for the EUV
and X-ray range into reach, with radiation properties which are so far only
available at large-scale facilities. 

Several different mechanisms of radiation generation at laser wakefield
accelerators (LWFA) are being  investigated and partially already used, both for
diagnosis  of the LWF accelerated electron bunches and for applications. The
most prominent examples are the generation of betatron radiation during the
acceleration process \cite{Rousse+PRL-93_2004, Phuoc+PhysPlasm-12_2005}, 
Thomson back scattering and inverse Compton scattering
\cite{Catravas+MeasSciTech-12_2001, Schwoerer+PRL-96_2006,
Phuoc+NaturePhot-6_2012} of
a laser pulse colliding with the LWF accelerated electron bunch and LWFA-driven
free electron lasers (FELs) \cite{Nakajima+NIMA-375_1996, Gruener+2007,
Schroeder+2006, Anania+2010}. Similar to the classical radiation sources, 
the radiation generated by these different mechanisms is
attractive for different classes of applications, particularly for time resolved
methods benefiting from the ultra-short radiation pulses which can be generated.
Betatron radiation as the
simplest of the above schemes is already being successfully applied in imaging
and spectroscopic applications \cite{Kneip+ApplPhysLett-99_2011,
Wenz+NatureComm-6_2015}. 

For given electron source
properties, particularly for given electron energy and \emph{average} beam current,
the free electron laser scheme is superior to all other schemes in terms of
spectral radiation power, (small) radiation energy bandwidth and coherence. It
is, therefore, the scheme of choice for a number of important classes of
applications like EUV lithography (integrated circuit fabrication), phase
contrast imaging (medical imaging) or single-molecule diffraction (life
sciences). Obviously, such applications would tremendously benefit from the
availability of laboratory-scale FELs which could be integrated into the
respective fabrication, hospital or laboratory infrastructure.

The FEL scheme, on the other hand, places the highest demands on the beam
quality provided by the accelerator. As it has been found,
\cite{Schroeder+FEL_2012}, laser wakefield accelerators can provide an initial normalized 6D beam
brightness comparable to that of conventional state-of-the-art accelerators,
which is a precondition for making them suited to drive an FEL. It is, however,
not straightforward to derive advantage from this initial condition for FEL
applications: particularly an energy spread on the percent level and an initial
beam divergence in a several milliradian range pose serious challenges to the
beam capture and transport and, if not counteracted, compromise the FEL
amplification. 

Therefore, to realize LWFA-driven FELs, an appropriate phase space manipulation
preparing the plasma-accelerated electron bunches for the FEL amplification
process is indispensable and a modification also of the FEL undulators might be
beneficial. Two schemes for driving FELs with large-energy spread beams have
been proposed: longitudinal decompression of the beam can be used to reduce the
\emph{slice} energy spread over the FEL cooperation length
\cite{Maier+2012,Couprie+JPhysB-47_2014} on the one hand and the scheme based on transverse gradient
undulators (TGUs) on the other hand, which will be discussed in the following.

The idea of the TGU scheme \cite{Smith+JApplPhys-50_1979,Fuchert+2012} is to
imprint a transverse spectral dispersion on
the beam, i.e. to correlate the particle energy with the particle's transverse
position. The amplitude of the magnetic flux density inside the TGU is a function of
the transverse position as well. If the transverse electron beam dispersion and the transverse magnetic
field distribution are properly matched the reference particles of all energies
present in the beam oscillate with the same frequency
 and radiate at the same wavelength
\begin{equation}
  \lambda = \frac{\lambda_\undu}{2\gamma^2(x)}\left(1+\frac{K_\undu^2}{2}\right)
  \label{eq:undulatorequation}
\end{equation}
with $\lambda_\undu$ the undulator period length, $\gamma$ the Lorentz factor, 
$K_\undu(x)=\frac{e}{m_ec}\lambda_\undu \tilde{B}_y(x)$ and $\tilde{B}_y$ the
magnetic flux density amplitude of the undulator. Thinking of
the electron bunch as being composed of a set of mono-energetic sub-bunches, it is
now the \emph{effective} energy spread introduced by the finite transverse
size of these sub-bunches together with the transverse magnetic field gradient
rather than the overall energy spread of the bunch which limits the FEL
amplification. Huang et al. \cite{Huang+2012} have shown that this limitation is
normally much less severe than that imposed by the energy spread and that with
the TGU scheme --- for the investigated examples --- a shorter
gain length and higher saturation power than without a TGU and even than with a
decompression scheme can be achieved, bringing compact LWFA-driven FELs into
reach.

In this contribution, we review the ongoing experimental projects towards TGU-based
LWFA-driven FELs and the results achieved in these projects so far.
\section{Experimental Projects towards LWFA-TGU-FELs}
Two collaborations are currently carrying out experimental projects
towards TGU-based LWFA-driven FELs: A consortium of the Chinese institutes SIOM
(Shanghai Institute of Optics and Fine Mechanics) and SINAP (Shanghai Institute
of Applied Physics) with the SLAC National
Accelerator Laboratory, US, on the one hand, and a consortium of the KIT, the
Helmholtz Institute Jena and the Friedrich-Schiller-University Jena in Germany,
on the other hand.

The SIOM/SINAP/SLAC TGU beam line is being set up at the SIOM
\SI{200}{\tera\watt} laser facility. In the SIOM LWFA set-up, a superposition of
two supersonic gas jets is used to control the plasma density profile and
thereby the internal injection and the wakefield acceleration process. In this
way electron bunches with an energy tunable between 200 and
\SI{600}{\mega\electronvolt}, energy spreads in the order of
\SI{1}{\percent} and a shot-to-shot energy jitter of $\pm\SI{5}{\percent}$ are
produced,
featuring bunch charges of up to \SI{80}{\pico\coulomb} and a beam divergence in the
order of \SI{0.3}{\milli\radian} \cite{Wang+PRL-117_2016}. For the TGU
experiment, a design beam energy of \SI{380}{\mega\electronvolt} has been chosen
\cite{Liu+PRAB-20_2017}.

For transporting the beam and matching the dispersion and beta functions to the
TGU, a beam line with a single deflection is foreseen. For the beam capture, a
strong quadrupole triplet, partly in vacuum, is foreseen, followed by the deflecting dipole and
a matching quadrupole triplet. To mitigate the increase of the projected normalized
emittance introduced by the energy spread of the beam as well as non-linear
dispersion effects generated by the strong quadrupoles, a correction scheme with three
sextupoles is used which requires sextupole strengths in the order of
\SI{e4}{\tesla\per\metre\squared} \cite{Liu+PRAB-20_2017}. These and the
in-vacuum quadrupoles are currently under technical development, whereas the
deflecting dipole and the second triplet together with a set of four permanent
magnet transverse gradient undulators are already in place. Each of these
undulators features 40 periods with a period length of \SI{20}{\milli\metre}, a
central deflection parameter $K_{\undu 0}=1.15$ and a relative transverse gradient
\begin{equation}
  \alpha := \frac{1}{K_{\undu 0}}\frac{\mathrm{d}K_\undu}{\mathrm{d}x}
  \label{eq:def_alpha}
\end{equation}
of \SI{50}{\per\metre} \cite{Liu+ProcIPAC_2016}. 


The Jena-Karlsruhe experiments have so far been performed at the JETI-40 laser
facility in Jena. Whereas the experimental setup at SIOM directly aims at a
TGU-FEL demonstration, our experimental approach is taking the intermediate step
of investigating the generation of spontaneous TGU radiation. The design energy
for the Jena-Karlsruhe set-up is \SI{120}{\mega\electronvolt}. A very large
energy acceptance of $\frac{\Delta E}{E}=\pm\SI{10}{\percent}$ is aimed at,
covering both, the single-bunch energy spread and the shot-to-shot energy
jitter.  The beam transport, as
discussed in more detail below, has an achromat-like dogleg lattice with two
deflecting dipoles, the TGU is a superconducting undulator with a cylindrical
pole shape.

\section{Progress on Experiments in the Jena-Karlsruhe Project}
\subsection{Superconducting TGU}
At KIT a 40-period superconducting transverse gradient undulator has been
realized and successfully tested in the vertical bath cryostat and field
measurement set-up CASPER. 

The design, realisation and test of this TGU from the very beginning aimed at
demonstrating the feasibility of the short period and high transverse field
gradient at a still sufficiently high $K_\undu$-parameter, which is required for
a compact and bright LWFA-driven EUV or X-ray source. 

\begin{figure}[t]
  \centering
  \includegraphics{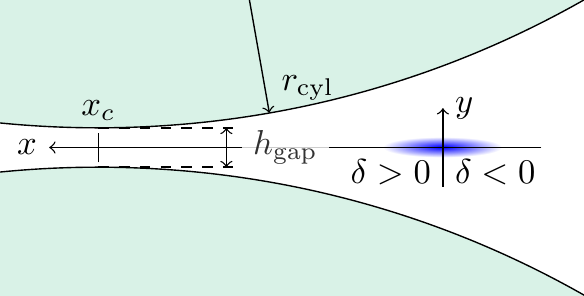}
  \caption{Pole geometry of the superconducting TGU. Additionally the transverse
  coordinate system used in the following is shown. The position of the
reference particle is $x=0, y=0$, $x_c$ is the $x$-position of the undulator
center, $h_\mathrm{gap}$ the gap height at this position, $\delta$ the relative
deviation of the particle energy from the reference energy.}
  \label{fig:tgu-geometry}
\end{figure}
\begin{table}
  \centering
  \begin{tabular}{lr}
	\toprule
	period length $\lambda_\undu$&\SI{10.5}{\milli\metre}\\
	gap @ symmetry axis $h_\mathrm{gap}$&\SI{1.1}{\milli\metre}\\
	pole radius $r_\mathrm{cyl}$&\SI{30}{\milli\metre}\\
	flux density ampl. $\tilde{B}_y(0)$&\SI{1.1}{\tesla}\\
	undulator parameter $K_{\undu 0}$ 	&	1.1\\
	transverse gradient $\frac{\partial K_\undu}{\partial x}$
	&\SI{149}{\per\metre}\\
	energy acceptance&$\pm\SI{10}{\percent}$\\
	\bottomrule
  \end{tabular}
  \caption{Main design parameters of the superconducting TGU}
  \label{tab:sctgu-parameters}
\end{table}
Particularly the demand for a high transverse gradient led to the choice of a cylindrical pole
shape as depicted in Fig.~\ref{fig:tgu-geometry}. Compared to a transverse
taper the cylindrical geometry has several advantages \cite{Fuchert+2012,
Afonso+ProcIPAC2011}. The achievable transverse
magnetic flux density gradient is much higher. The magnetic flux density profile
exhibits a flexion point around which the field gradient is constant to a very
good approximation. Lastly, this geometry is comparably simple to realize,
particularly in case of superconducting undulators. The main parameters of the
SCTGU resulting from the magnetic design optimization \cite{Afonso+2012,
Afonso2015} are summarized in
Table~\ref{tab:sctgu-parameters}.

Due to the transverse gradient in the magnetic flux density the particles
traveling through the TGU experience a net kick towards the lower-field region
in each period which leads to a transverse drift. This ponderomotive effect can
be suppressed by superimposing a $x$-dependent correction field constant in
longitudinal direction. In case of the SCTGU, this correction field is a weak
($\sim\SI{0.5}{\milli\tesla}$) combined dipole and shifted sextupole field which
is generated by two long, narrow racetrack coils which are inserted in the coil
formers. In order to avoid distortions of this correction field, the undulator
is entirely iron-free.

\begin{figure}
  \centering
  \includegraphics[width=.9\columnwidth]{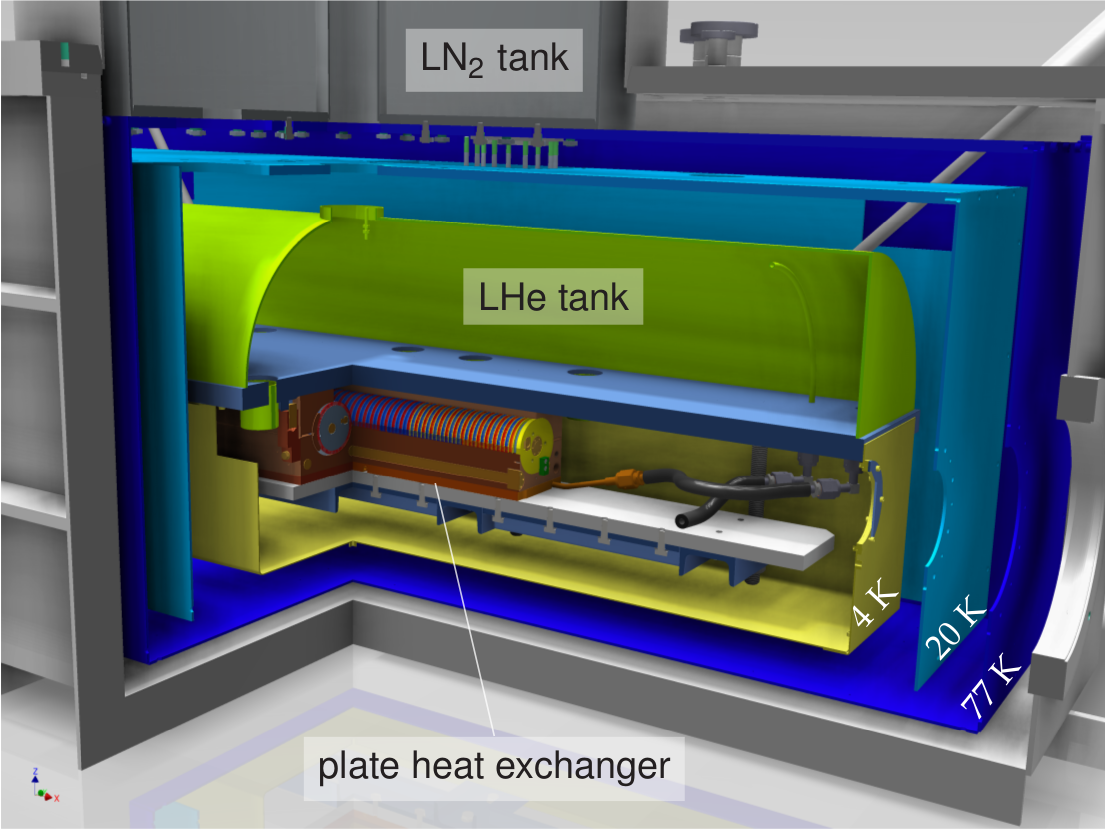}
  \caption{Cutaway view of the SCTGU cryostat.}
  \label{fig:cryostat}
\end{figure}
Both the undulator coils and the correction coils are wound from Nb-Ti
multifilament wires. For operating the undulator at the JETI beam
line a specialized cryostat was constructed. A cutaway view of the cryostat
is shown in Fig.~\ref{fig:cryostat}. The undulator is placed in the beam vacuum
and indirectly cooled through plate heat exchangers with a liquid Helium flow
from a reservoir placed on top of the undulator.

\begin{figure}[tb]
  \centering
  \includegraphics{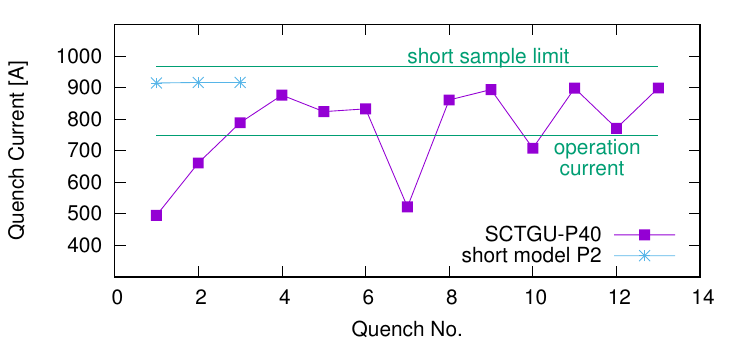}
  \caption{Quench history of the SCTGU short model and of the full-scale
    SCTGU-P40.
  The short sample quench limit of the Nb-Ti multifilament wire and the foreseen
operation current are indicated by horizontal lines.}
  \label{fig:sctgu-quenchtest}
\end{figure}
The SCTGU was realized in-house in three steps: first the general winding geometry and
the quench performance of single 2-period short model test coils were tested and
improved, second a complete 2-period short model was built, tested and
characterized and eventually the full-scale device was realized, tested and
partially characterized in the bath cryostat CASPER.
Figure~\ref{fig:sctgu-quenchtest} shows the quench history of the short model P2 and
the full-scale SCTGU-P40. The short model immediately reached a stable quench
current close to the short sample limit of the superconducting wire. The
SCTGU-P40, after 13 quenches has not yet reached entirely stable conditions.
However, the quench history indicates that it will be safely operable at the
design operation current of \SI{750}{\ampere}.

\begin{figure}[tb]
  \centering
  \includegraphics{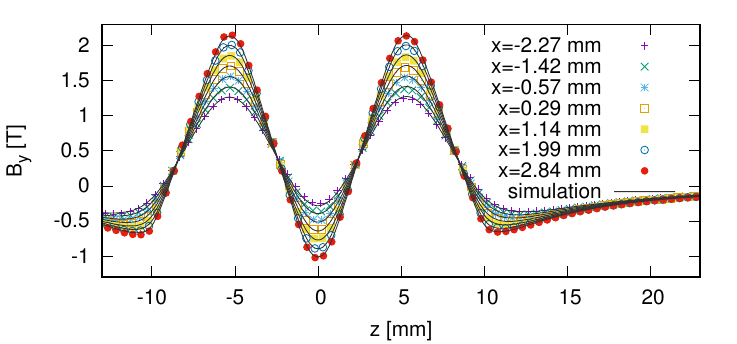}\\
  \includegraphics{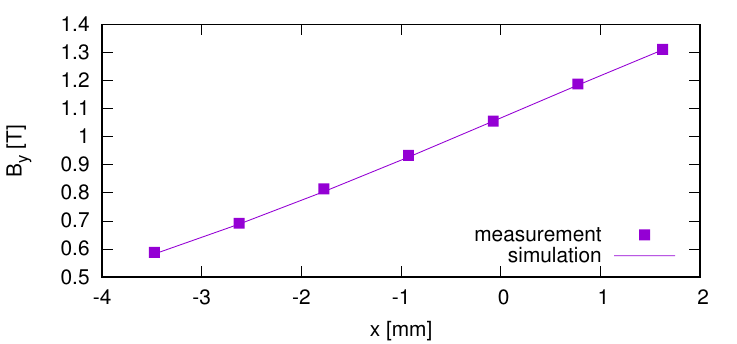}
  \caption{Results of the magnetic flux density measurements, top: SCTGU short
  model, $z$-scan with an array of seven Hall probes at equidistant
  $x$-positions, shifted from the median plane to $y=\SI{0.6}{\milli\metre}$ due
to the finite thickness of the Hall probe array; bottom: field measurement with
the same Hall probe array at the fixed $z$-position of one pole of the central
period of the 40-period SCTGU. The estimated accuracy of the measurement in both
cases is \SI{1}{\percent}. The measurements are compared to simulations with
OPERA 3D which take into account the positioning of the Hall probe array.}
  \label{fig:sctgu-fieldmeasurement}
\end{figure}
For the magnetic characterization of both devices a Hall probe array supplied by
AREPOC s.r.o., Slovakia, was employed, measuring the vertical ($y$-) component of
the magnetic flux density at seven equidistant transverse ($x$-) positions
around the beam position in the TGU gap. In case of the short model, this array
could be scanned through the undulator gap in longitudinal ($z$-) direction. Due to
geometrical limitations of the set-up in CASPER, a field measurement at only
one fixed $z$ position was performed. The results of these measurements, shown
in Fig.~\ref{fig:sctgu-fieldmeasurement}, show excellent agreement with the
theoretical predictions. Free Electron Lasers (FEL) are commonly regarded as the potential key
 application of laser wakefield accelerators (LWFA). It has been found that electron
 bunches exiting from state-of-the-art LWFAs exhibit a
 normalized 6-dimensional beam brightness comparable to those in
 conventional linear accelerators. Effectively exploiting this beneficial beam property for
 LWFA-based FELs is challenging due to the extreme initial conditions
 particularly in terms of beam divergence and energy spread. Several different
 approaches for capturing, reshaping and matching LWFA beams to suited
 undulators, such as bunch decompression or transverse-gradient undulator
 schemes, are currently being explored. In
 this article the transverse gradient undulator concept will be discussed with a focus on recent
 experimental achievements.

\subsection{Beam transport}
\subsubsection{Matching conditions and beam optics}
\begin{figure}[tb]
  \centering
  \includegraphics[width=7.5cm]{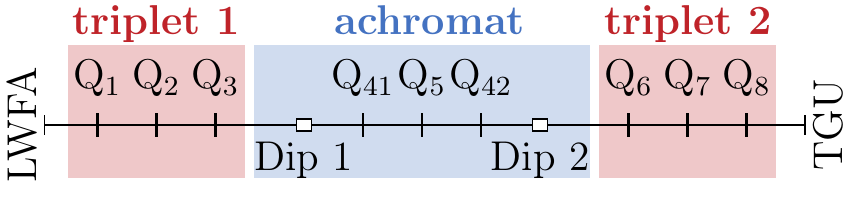}
  \caption{Schematic layout of the magnetic lattice of the beam transport line
  from the LWFA to the TGU, consisting of a matching quadrupole triplet, an
achromat-like cell with two dipoles and three quadrupoles, and a second matching
quadrupole triplet.}
  \label{fig:chicane-lattice}
\end{figure}
Our beam transport line design started off with the (geometrically) symmetric
magnet lattice schematically shown in Fig.~\ref{fig:chicane-lattice}, consisting
of a matching quadrupole triplet close to the LWFA, an achromat-like
dogleg chicane and a second matching triplet at the entrance of the TGU
\cite{Widmann+2014, Widmann2016}. 

Approximative matching conditions for the TGU can be derived from the undulator
equation as the resonance condition and a linear TGU
model which yields
\begin{equation}
  D = \frac{2+K_{\undu 0}^2}{\alpha K_{\undu 0}^2}
  \label{eq:tgu-dispersionmatching}
\end{equation}
for the linear dispersion \cite{Huang+2012, Baxevanis+2014} and
\begin{equation}
  \begin{aligned}
  \sqrt{\beta_x\epsilon_x}&<
  \frac{1+\frac{K_\undu^2}{2}}{K_\undu^2\alpha}\frac{\sigma_\lambda}{\lambda} \\
  \frac{\epsilon_x}{\beta_x}&< 
  \frac{1+\frac{K_\undu^2}{2}}{\gamma^2}\frac{\sigma_\lambda}{\lambda}
  \end{aligned}
  \label{eq:tgu-betamatching}
\end{equation}
for the beam size and divergence, respectively, inside the undulator
\cite{Bernhard+PRAB-19_2016}. Here,
$\epsilon_x$ is the geometric emittance of a monochromatic slice of
the electron beam, $\frac{\sigma_\lambda}{\lambda}$ is to be understood as the
relative wavelength bandwidth condition which in case of spontaneous undulator
radiation scales like $\frac{1}{N_\undu}$ with $N_\undu$ the number of
undulator periods, and in an idealizing case of FEL amplification like the Pierce
parameter with $\frac{3\pi}{2}\rho_\mathrm{FEL}$. In our case, these estimations
lead to the condition that the slice transverse beam size in the deflection direction
should be clearly below \SI{e-4}{\metre} inside the undulator. 

It is worth noting that a TGU due to the transverse field gradient with
periodically alternating field direction and therefore also alternating sign of
the gradient forms a periodically focussing and defocussing (FODO) structure. 
Thus there exists a periodic solution for
$\beta_x$ inside the undulator. For the SCTGU operated at the design field amplitude
and with the design beam energy of \SI{120}{\mega\electronvolt} the amplitude of
the periodic betatron function turns out to be too large
($\tilde\beta_x=\SI{2.2}{\metre}$) to fulfill the above beam size condition.
Consequently, a matching scheme with an external focusing to the undulator
center was chosen for $\beta_x$. For the vertical direction, a matching to the
constant beta-function solution is favorable, which exists due to the natural
focusing present in any planar insertion device. 

\begin{figure}[tb]
  \centering
  \includegraphics{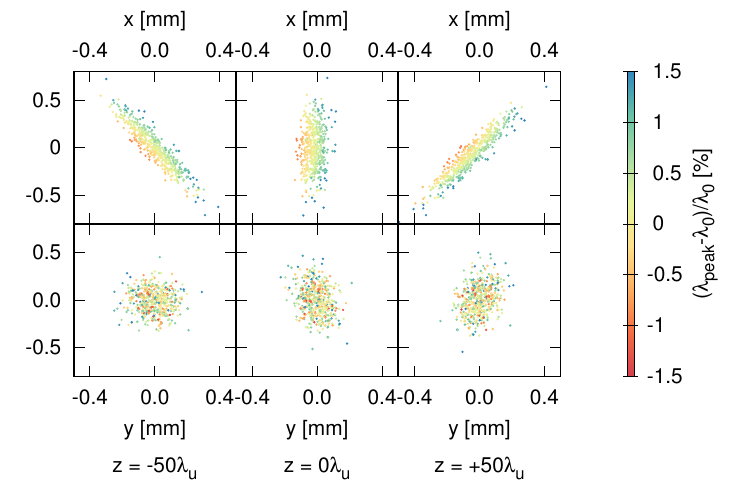}
  \caption{Phase space distributions of a bunch with energy
    $E=\SI{120}{\mega\electronvolt}$ and gemetric slice emittance $\epsilon_{x,y}=\SI{e-8}{\meter\radian}$, 
    focused into the TGU in $x$ with $\beta_{x w}=\SI{0.5}{\meter}$ and matched to
    the constant beta function $\beta_y=\SI{0.7}{\meter}$ in $y$. The
  distributions at the entrance, the center and the exit of a 100-period TGU
with the parameters of our SCTGU. The color code describes the relative
wavelength shif of the radiation emitted by the respective particle with respect
to the reference wavelength.}
  \label{fig:tgu-phasespace}
\end{figure}
\begin{figure}[tb]
  \centering
  \includegraphics{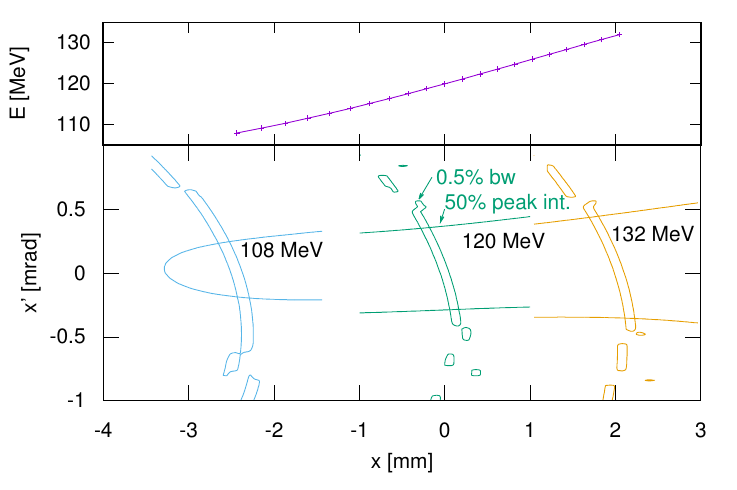}
  \caption{Matching conditions at the entrance of the SCTGU, top: required
  dispersion, bottom: acceptance contours in $x-x'$ phase space for three beam
  slice energies $E=\SI{120}{\mega\electronvolt}\pm\SI{10}{\percent}$, as
defined by monochromaticity and intensity conditions on the radiation field}
  \label{fig:tgu-completematching}
\end{figure}
For a better understanding of the matching conditions for finite-emittance beams
we investigated the spontaneous radiation fields of the particles of a finite-emittance
bunch tracked through the undulator \cite{Bernhard+PRAB-19_2016}. Figure~\ref{fig:tgu-phasespace} shows the
phase space distributions in $x$ and $y$ at the entrance, center and exit of the
TGU resulting from a simulation with WAVE \cite{Scheer2012} for a 100-period TGU with
our design parameters. The simulation was done for a (slice) particle beam at the
design beam energy $E_0=\SI{120}{\mega\electronvolt}$ with a slice geometric
emittance $\epsilon_{x,y}=\SI{10}{\nano\metre\radian}$,
$\beta_y=\SI{0.7}{\metre}=\mathrm{const.}$ and focused in $x$ to the TGU center
with $\beta_{x w}=\SI{0.5}{\metre}$. The color code reflects the relative
deviation of the wavelength of the radiation emitted by the respective particle
from the design wavelength $\lambda_0$. This result suggests that there is
virtually no correlation between the wavelength detuning and the position in the
$y-y'$ phase space, whereas the strong correlation between detuning and position
in $x-x'$ phase space is dominated by the effect of the finite beam size in the
beam waist, while the detuning due to the beam divergence appears much weaker:
low $\beta_{x w}$ values are favoured. Such tracking simulations performed for
several beam slices at discrete beam energies may be used to
establish overall matching conditions for the TGU in terms of optimized
dispersion conditions and acceptance areas in $x-x'$ phase space.
Figure~\ref{fig:tgu-completematching} shows the result for the 
Jena-Karlsruhe TGU extended to 100 periods and beam energies of
$\SI{120}{\mega\electronvolt}\pm\SI{10}{\percent}$. 

\begin{table}
  \centering
  \begin{tabular}{lcc}
	  \toprule
	  &initial&final\\
	  \midrule
	  $E_0$&\multicolumn{2}{c}{\SI{120}{\mega\electronvolt}}\\
	  $\epsilon_{x, y}$&\multicolumn{2}{c}{\SI{10}{\nano\metre\radian}}\\
	  $\beta_{x}$&\SI{1.6e-3}{\metre}&\SI{1.6}{\metre}\\
	  $\alpha_{x}$&0&2.6\\
	  $\beta_{y}$&\SI{1.6e-3}{\metre}&\SI{0.7}{\metre}\\
	  $\alpha_{y}$&0&0\\
	  \bottomrule
  \end{tabular}
  \caption{Matching conditions for the beam transport line at the LWFA (initial)
  and the TGU (final) in terms of linear optical functions.}
  \label{tab:matching-conditions}
\end{table}
\begin{figure}[tb]
  \centering
  \includegraphics{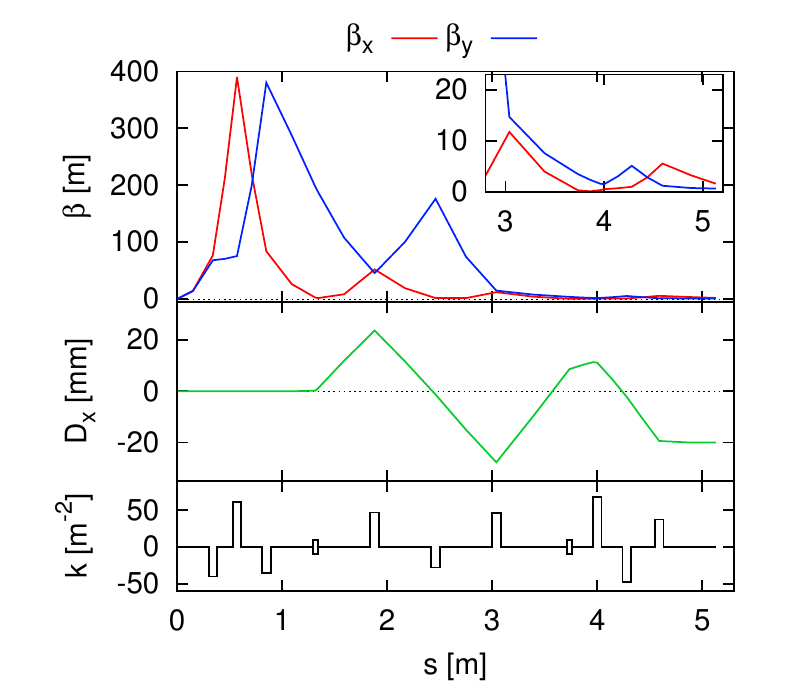}
  \caption{Optical functions of the beam transport line matched to the
    initial and final conditions summarized in
    Table~\ref{tab:matching-conditions}}
  \label{fig:achromat-opticalfunctions}
\end{figure}
Table~\ref{tab:matching-conditions} summarizes the matching conditions at the
exit of the LWFA (initial), estimated from the plasma wake structure size and
the beam divergence measurements, and at the entrance of the TGU (final) as
derived from the above considerations. The initial and final triplet of the beam
transport lattice can be matched to the initial and final conditions, and in a
second step the dispersion is matched through an adjustment of the achromat cell
and the final triplet. The result of this matching is shown in
Fig.~\ref{fig:achromat-opticalfunctions} for the design beam energy
\SI{120}{\mega\electronvolt}. 

Clearly, to make this matching work for a
sufficiently wide band of electron energies, higher order corrections are
necessary. For the final realization stage we foresee the application of
combined-function quadrupole-sextupole magnets at the positions of $Q_{41}$,
$Q_{42}$, $Q_{6}$, $Q_{8}$. 

We note that the shown optics without being optimized for bunch length
preservation leads to only a moderate bunch lengthening with the transport
matrix elements $R_{51}=0.1$, $R_{52}=\SI{2e-3}{\metre}$ and
$R_{56}=\SI{-4e-4}{\metre}$. There is probably the potential to optimize the
lattice towards a close to isochronous beam transport.

\subsubsection{Experimental realization}
In the experimental realization of our beam transport concept, we started with a
linear beam transport line consisting of the first quadrupole triplet and the
achromat-like dogleg chicane cell. The idea is to investigate the TGU radiation
for a range of electron energies by consecutively adjusting the focusing for
each energy while keeping the dispersion fixed. For this approach fewer degrees
of freedom are required than in the fully matched and chromatically corrected
case, allowing us to omit the second matching triplet.

The beam transport line was set up at the JETI-40 laser facility in Jena, at
that time providing laser pulses with a pulse duration of
\SI{28}{\femto\second}, a pulse energy of up to \SI{1.2}{\joule} before
compression, and a peak intensity at the target of
\SI{9.1e18}{\watt\per\centi\metre\squared}. For
the plasma accelerator, a gas cell
($\varnothing\SI{1}{\milli\metre}\times\SI{3}{\milli\metre}$) with a
\SI{95}{\percent} He and \SI{5}{\percent} N$_2$ gas mixture was used.

To allow for maximum experimental flexibility, the complete beam line is realized
with electromagnets placed in an about 3-m long vacuum box connected to
the vacuum chamber of the LWFA. The chicane dipoles were supplied by GMW
Associates, the in-vacuum quadrupoles and later on combined function magnets are
in-house designed and manufactured. The philosophy behind that is again
experimental flexibility: the magnets are comparably simple, inexpensive and
thanks to a modular design easy to modify. In the current set-up, two types of
quadrupoles with a gap radius \SI{11}{\milli\metre}, a magnetic length of
\SI{80}{\milli\metre} and a nominal field gradient of \SI{30}{\tesla\per\metre}
and \SI{39}{\tesla\per\metre}, respectively, were used
\cite{Bernhard+ProcIPAC_2015a}. 

All magnets were characterized by 3D Hall probe scans at our magnetic
measurement set-up \cite{Hillenbrand+ProcIPAC_2015}. Besides a classical 3D mapping
of the three Cartesian components of the magnetic flux density the method of circular
Hall probe scans was applied which yields a 1D map of the multipole components
of the magnetic field along the axis. This mapping can very effectively be used
in tracking simulations taking into account the fringe fields and multipole
errors of the real magnets to a good approximation \cite{Will+ProcIPAC_2017}. We used
this representation in comparison to a simple hard edge model in our tracking
simulations in order to disentangle general properties of the beam transport
optics from effects due to individual properties of the non-perfect magnets.

A LANEX screen insertable at three longitudinal positions --- after the first
triplet, after the central quadrupole of the achromat cell and after the final
dipole --- and an electron spectrometer taking the place of the undulator served
as electron beam diagnostics in this experiment.

\subsubsection{Experimental results}
\begin{figure}[tb]
  \centering
  \includegraphics{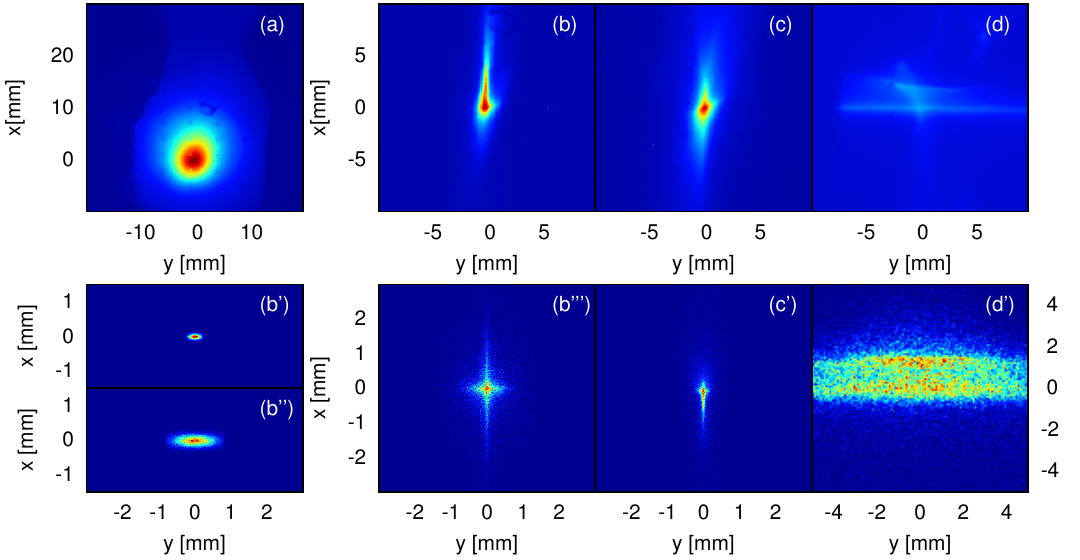}
  \caption{Beam images taken during the experimental test of the beam transport
  line (a--d) and results of tracking simulations (b'--c'). The color coding
corresponds to the collected charge in arbitrary units (not equally scaled).
Refer to the text for explanations.}
  \label{fig:beamimages}
\end{figure}
In the experiments on the beam transport were performed at electron energies in
the range of \SIrange{20}{70}{\mega\electronvolt} due to limitations of the
laser system present at that time. An image of the initial beam at the LANEX
screen positioned \SI{0.7}{\metre} after the gas cell, averaged over 180
consecutive shots, is shown in Fig.~\ref{fig:beamimages} (a). This image is
resulting from the convolution of the divergence of the individual bunches ---
$\SI{9.5}{\milli\radian}\times\SI{10.5}{\milli\radian}$ (FWHM) on average ---
and a pointing jitter of $\SI{2.4}{\milli\radian}\times\SI{4.2}{\milli\radian}$
(rms). The asymmetry in the vertical profile originates from the latter. This
rather unstable pointing implied that averaging over about 30 consecutive shots
was necessary in each step of the experiment. The asymmetry of the initial
averaged profile is visible also in the averaged profiles of the captured and transported
beam.

\begin{figure}[tb]
  \centering
  \includegraphics[width=7.5cm]{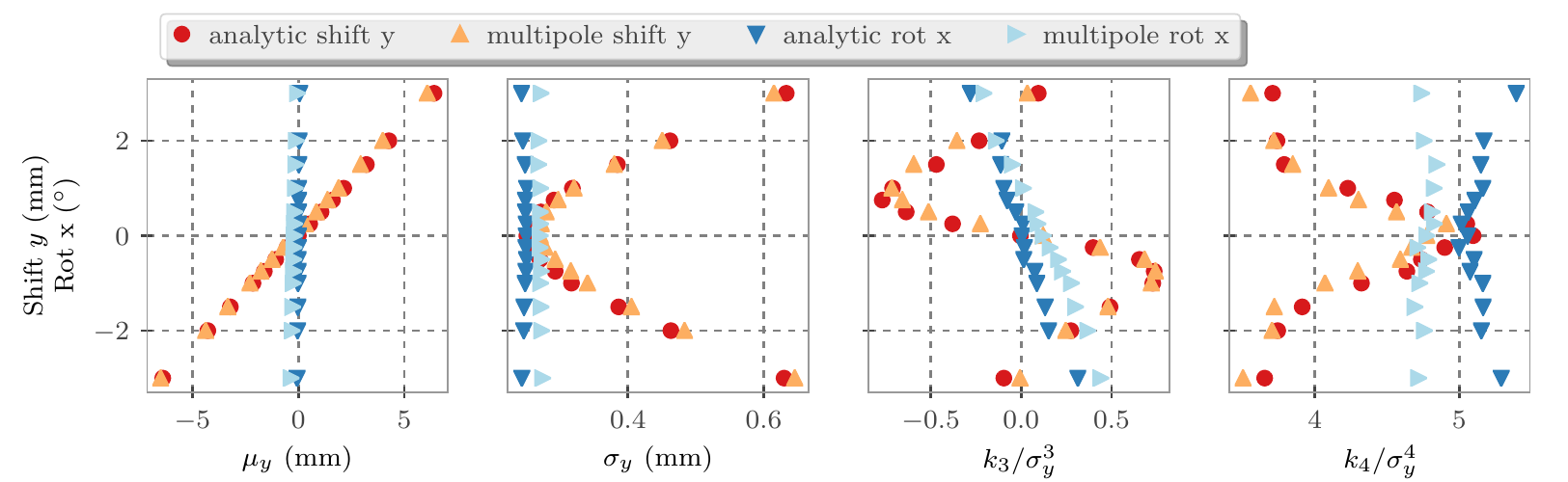}
  \caption{First four statistical moments of the beam profile in focusing
  direction observed on a beam screen after one quadrupole, as a function of
misalignment of the quadrupole for a translational misalignment in focusing
direction and a rotational misalignment around the transverse axis perpendicular
to the focusing direction. The moments have been calculated from simulated
screen images, based on tracking of a Gaussian beam. Tracking results with two
different representations of the magnetic field are compared, a hard edge model
(analytic) and a 1D multipole map retrieved from the magnetic characterization
of the quadrupole (multipole).}
  \label{fig:statmoments}
\end{figure}
Since also the mean pointing direction of the electron bunches differed slightly
from that of the laser to which the magnets were pre-aligned, the first step of
the experiment was to adjust the magnet alignment with respect to the mean beam
axis. The transverse position of each individual quadrupole was remotely
adjustable in two (first triplet) or, respectively, one (second triplet)
translational degrees of freedom. If the magnet is turned on and scanned in its
focusing direction, the profile of the focused beam captured on a subsequent
screen can be used for a beam-based alignment of the magnet. Tracking
simulations show that the skewness of the beam profile in focused direction is
particularly sensitive to quadrupole misalignments, both translational about the
axis perpendicular and rotational about the transverse axis parallel to the
focusing plane. The variance and kurtosis of the profile are sensitive to
translational alignment errors only. These characteristics shown in
Fig.~\ref{fig:statmoments} can in principle be utilized for a successive
beam-based alignment of the magnet in the various transverse degrees of freedom.
Also shown in the figure is a comparison of tracking results using the above mentioned 1D multipole representation of the magnet
deduced from the circular Hall probe scans with those using an analytic
hard-edge model. The influence of the fringe fields and multipole errors of the
real magnets on the beam profile characteristics is visible, but small.

\begin{figure}[tb]
  \centering
  \includegraphics{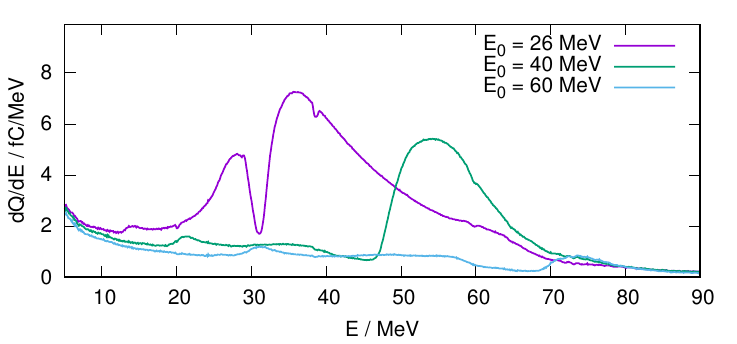}\\
  \includegraphics[width=7.5cm]{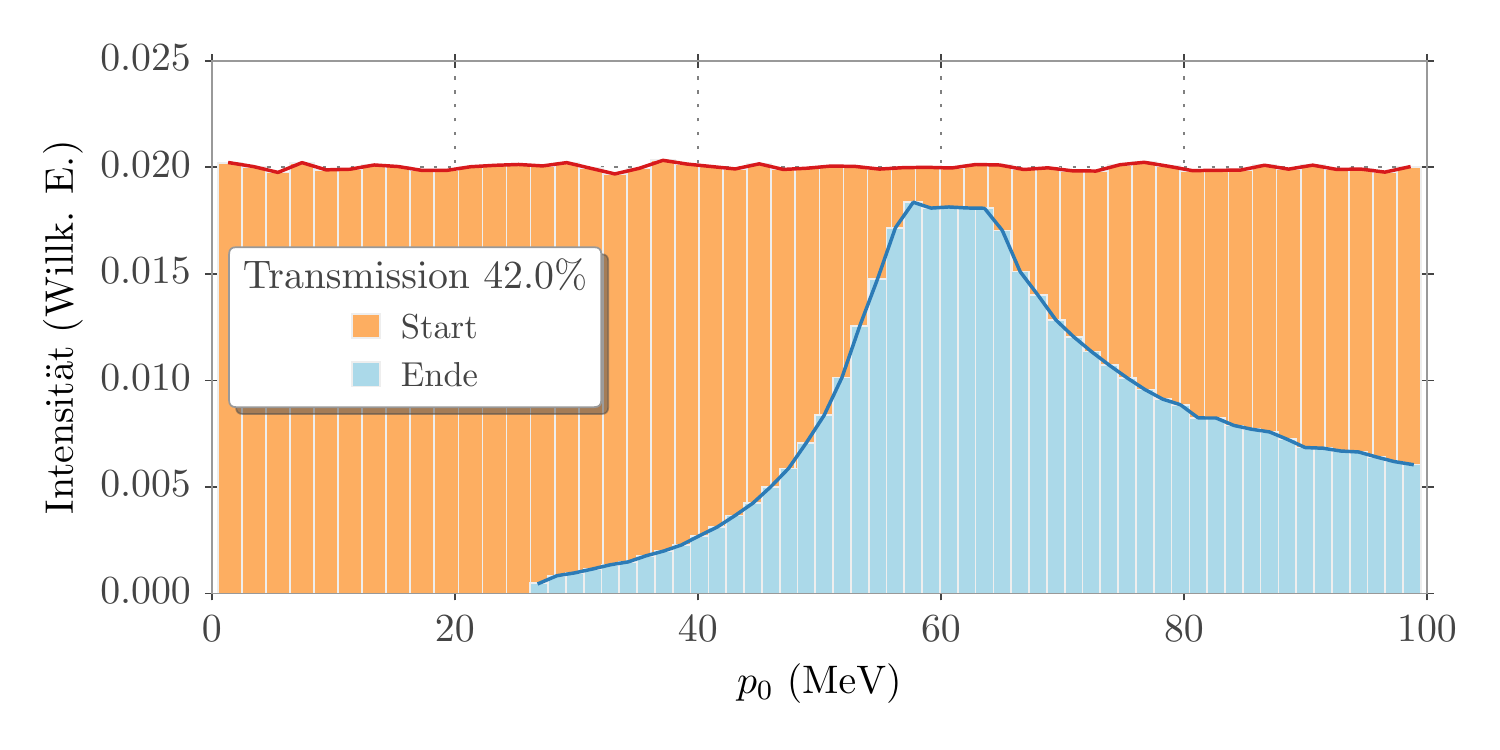}
  \caption{Electron spectra measured in the electron spectrometer
    \SI{3.3}{\metre} downstream from the gas cell for three different settings
  of the first matching triplet (top). Spectral transmission function of the
  first triplet adjusted for focusing a \SI{60}{\mega\electronvolt} beam into
the electron spectrometer, calculated by tracking simulations.}
  \label{fig:spectra}
\end{figure}
After the beam-based alignment of the first triplet it was possible to focus the
electron bunches into the electron spectrometer \SI{3.3}{\metre} downstream from
the gas cell. The electron energy spectra measured with focusing settings for
three different energies are shown in Fig.~\ref{fig:spectra}. For comparison,
the spectral transmission function for the triplet setting for
$E_0=\SI{60}{\mega\electronvolt}$ deduced from tracking simulations is shown. 

Figure~\ref{fig:beamimages} (b)--(d) shows the averaged electron beam images at
different beam screen positions taken at the further steps of
setting up the beam transport to the TGU: (b) using the first triplet focusing
to the screen \SI{1.9}{\metre} downstream from the gas cell, (c) using the first
triplet focusing to the screen \SI{3.2}{\metre} downstream with the two chicane
dipoles on and (d) using additionally the quadrupoles of the achromat cell and
with the setting satisfying the matching conditions for the TGU (at
\SI{40}{\mega\electronvolt}) --- note that in
this case the beam focus is about \SI{1}{\metre} behind the beam screen.

Figure~\ref{fig:beamimages} (b')--(d') show for comparison the results of
tracking simulations investigating the influence of the initial beam properties
on beam profiles which observable for the respective optics settings. These
simulations assume perfect magnets and magnet alignments and a Gaussian initial
beam. Such simulations give an idea how the initial beam properties in terms of
divergence and energy spread contribute to the features observed. In
Fig.\ref{fig:beamimages} (b') and (b'{}') the images are shown that would be
expected if the beam had a negligible energy spread. The large divergence of the
initial beam has mainly the effect of increasing the spot size without affecting
the Gaussian spot shape (b'{}'), as the
comparison to a beam with an initial beam divergence of \SI{1}{\milli\radian}
(b') shows. An additional increase of the energy spread (b'{}'{}') leads to the
observed cross shape. Fig.~\ref{fig:beamimages} (c') and (d') do not represent a
rigorous theoretical reproduction of the experimental results. 
These calculations guide our general understanding of the obtained
experimental results and, more important, establish a starting point for
preparing the next experimental steps and related modelling. 
\subsection{Next steps}
Obviously, the next step is to set up the complete proof-of-principle experiment
encompassing beam transport, transverse gradient undulator and photon
diagnostics. The work on two major preparatory tasks is in progress. Firstly,
the cryostat for the SCTGU has recently successfully been commissioned; a
sliding system for scanning the above mentioned Hall probe array has been built
and tested and will be installed along with the TGU in the cryostat for a full
2D magnetic characterization of the TGU. 

Secondly, an improved positioning system for the quadrupoles is under
construction which will provide a much better control over the alignment of the
magnets and add the rotational and translational degrees of freedom which were
missing in the above-described experiments. 
\section{Conclusion}
Scaling considerations and simulations suggest that transverse-gradient
undulator schemes are a valid option for realizing compact, laser wakefield
accelerator-driven FELs. An experimental validation of the TGU concept is still
due. The two major projects towards this experimental proof of principle
conducted by SIOM/SINAP/SLAC on the one hand and FSU Jena/HI Jena/KIT on the
other hand, follow different conceptual, experimental and technological
approaches. On the side of the Jena-Karlsruhe project, major experimental
achievements were the demonstration of the technical feasibility of
high-transverse-gradient superconducting TGU and the successful performance of a
pioneering experiment on beam capture, transport and matching at the LWFA at the
JETI-40 laser facility. The electron beam generated by the LWFA could be
controlled in a quite robust manner and transported over \SI{3.5}{\metre} to the
later position of the TGU. The beam transport line set-up deliberately was kept simple and flexible
which turned out to be very useful in the experiment. However, due to the
limited diagnostics available a thorough characterization and good control of
the beam line components as well as accompanying simulations are essential for
this kind of experiment.

So far, the experimental results indicate that the goal of a TGU radiation source driven by
a LWFA is achievable.
\section{Acknowledgement}
The work described in this article has partially been supported by the German Federal Ministry of Education and
Research under Grant 05K10VK2 and Grant 05K10SJ2.

\bibliographystyle{unsrt}
\bibliography{physik}

\begin{thebibliography}{10}

\bibitem{Rousse+PRL-93_2004}
Antoine Rousse, Kim~Ta Phuoc, Rahul Shah, Alexander Pukhov, Eric Lefebvre,
  Victor Malka, Sergey Kiselev, Fr\'ederic Burgy, Jean-Philippe Rousseau,
  Donald Umstadter, and Dani\'ele Hulin.
\newblock Production of a kev x-ray beam from synchrotron radiation in
  relativistic laser-plasma interaction.
\newblock {\em Phys. Rev. Lett.}, 93:135005, Sep 2004.

\bibitem{Phuoc+PhysPlasm-12_2005}
Kim~Ta Phuoc, Fréderic Burgy, Jean-Philippe Rousseau, Victor Malka, Antoine
  Rousse, Rahul Shah, Donald Umstadter, Alexander Pukhov, and Sergei Kiselev.
\newblock Laser based synchrotron radiation.
\newblock {\em Physics of Plasmas}, 12(2):023101, 2005.

\bibitem{Catravas+MeasSciTech-12_2001}
P~Catravas, E~Esarey, and W~P Leemans.
\newblock Femtosecond x-rays from thomson scattering using laser wakefield
  accelerators.
\newblock {\em Measurement Science and Technology}, 12(11):1828, 2001.

\bibitem{Schwoerer+PRL-96_2006}
H.~Schwoerer, B.~Liesfeld, H.-P. Schlenvoigt, K.-U. Amthor, and R.~Sauerbrey.
\newblock Thomson-backscattered x rays from laser-accelerated electrons.
\newblock {\em Phys. Rev. Lett.}, 96:014802, Jan 2006.

\bibitem{Phuoc+NaturePhot-6_2012}
Phuoc~K. T., S.~Corde, C.~Thaury, V.~Malka, A.~Tafzi, J.P. Goddet, R.C. Shah,
  S.~Sebban, and A.~Rousse.
\newblock All-optical compton gamma-ray source.
\newblock {\em Nature Photonics}, 6, 2012.

\bibitem{Nakajima+NIMA-375_1996}
K~Nakajima, M~Kando, T~Kawakubo, T~Nakanishi, and A~Ogata.
\newblock A table-top x-ray fel based on the laser wakefield
  accelerator-undulator system.
\newblock {\em Nuclear Instruments and Methods in Physics Research Section A:
  Accelerators, Spectrometers, Detectors and Associated Equipment}, 375(1):593
  -- 596, 1996.
\newblock Proceedings of the 17th International Free Electron Laser Conference.

\bibitem{Gruener+2007}
F.~Gr\"{u}ner, S.~Becker, U.~Schramm, E.~Eichner, M.~Fuchs, R.~Weingartner,
  D.~Habs, J.~Meyer-Ter-Vehn, M.~Geissler, M.~Ferrario, L.~Serafini, B.~van~der
  Geer, H.~Backe, and W.~Lauth.
\newblock {Design considerations for table-top, laser-based VUV and X-ray free
  electron lasers}.
\newblock {\em Applied Physics B}, 86:431--435, 2007.

\bibitem{Schroeder+2006}
C.~B. Schroeder, W.~M. Fawley, E.~Esarey, and W.~P. Leemans.
\newblock {Design of an XUV FEL Driven by the Laser-Plasma-Accelerator at the
  LBNL LOASIS Facility}.
\newblock http://escholarship.org/uc/item/1sp9h9w2, 2006.

\bibitem{Anania+2010}
M.~P. Anania, E.~Brunetti, S.~Cipccia, D.~Clark, R.~Issac, G.G. Manahan,
  T.~McCanny, A.~J.~W. Reitsma, R.~P. Shanks, G.H. Welsh, S.~M. Wiggins, D.~A.
  Jaroszynski, S.B. van~der Geer, M.J. de~Loos, M.W. Poole, J.~Clarke, and
  B.~Shepherd.
\newblock {The ALPHA-X Beam Line: Toward a Compact FEL}.
\newblock In {\em {Proceedings of the International Particle Accelerator
  Conference IPAC2010}}, page TUPE052, 2010.

\bibitem{Kneip+ApplPhysLett-99_2011}
S.~Kneip, C.~McGuffey, F.~Dollar, M.~S. Bloom, V.~Chvykov, G.~Kalintchenko,
  K.~Krushelnick, A.~Maksimchuk, S.~P.~D. Mangles, T.~Matsuoka, Z.~Najmudin,
  C.~A.~J. Palmer, J.~Schreiber, W.~Schumaker, A.~G.~R. Thomas, and
  V.~Yanovsky.
\newblock X-ray phase contrast imaging of biological specimens with femtosecond
  pulses of betatron radiation from a compact laser plasma wakefield
  accelerator.
\newblock {\em Applied Physics Letters}, 99(9):093701, 2011.

\bibitem{Wenz+NatureComm-6_2015}
J.~Wenz, S.~Schleede, K.~Khrennikov, M.~Bech, P.~Thibault, M.~Heigoldt,
  F.~Pfeiffer, and S.~Karsch.
\newblock Quantitative x-ray phase-contrast microtomography from a compact
  laser-driven betatron source.
\newblock {\em Nature Communications}, 6, 2015.

\bibitem{Schroeder+FEL_2012}
C.~B. Schroeder, C.~Benedetti, E.~Esarey, W.~P. Leemans, J.~van Tilborg,
  Y.~Ding, Z.~Huang, F.~Gr\"{u}ner, and A.~Maier.
\newblock Application of laser-plasma accelerator beams to free-electron
  lasers.
\newblock In {\em Proceedings of the FEL2012, Nara, Japan}, 2012.

\bibitem{Maier+2012}
A.~R. Maier, A.~Meseck, S.~Reiche, C.~B. Schroeder, T.~Seggebrock, and
  F.~Gr\"{u}ner.
\newblock {Demonstration Scheme for a Laser-Plasma-Driven Free-Electron Laser}.
\newblock {\em Phys. Rev. X}, 2:031019, Sep 2012.

\bibitem{Couprie+JPhysB-47_2014}
M~E Couprie, A~Loulergue, M~Labat, R~Lehe, and V~Malka.
\newblock Towards a free electron laser based on laser plasma accelerators.
\newblock {\em Journal of Physics B: Atomic, Molecular and Optical Physics},
  47(23):234001, 2014.

\bibitem{Smith+JApplPhys-50_1979}
T.~I. Smith, J.~M.~J. Madey, L.~R. Elias, and D.~A.~G. Deacon.
\newblock {Reducing the sensitivity of a freeelectron laser to electron energy
  }.
\newblock {\em Journal of Applied Physics}, 50:4580, 1979.

\bibitem{Fuchert+2012}
Golo Fuchert, Axel Bernhard, Sandra Ehlers, Peter Peiffer, Robert Rossmanith,
  and Tilo Baumbach.
\newblock {A novel undulator concept for electron beams with a very large
  energy spread}.
\newblock {\em {Nuclear Instruments and Methods in Physics Research A}},
  672:33--37, 2012.

\bibitem{Huang+2012}
Zhirong Huang, Yuantao Ding, and Carl~B. Schroeder.
\newblock {Compact X-ray Free-Electron-Laser from a Laser-Plasma Accelerator
  Using a Transverse-Gradient Undulator}.
\newblock {\em PRL}, 109:204801, 2012.

\bibitem{Wang+PRL-117_2016}
W.~T. Wang, W.~T. Li, J.~S. Liu, Z.~J. Zhang, R.~Qi, C.~H. Yu, J.~Q. Liu,
  M.~Fang, Z.~Y. Qin, C.~Wang, Y.~Xu, F.~X. Wu, Y.~X. Leng, R.~X. Li, and Z.~Z.
  Xu.
\newblock High-brightness high-energy electron beams from a laser wakefield
  accelerator via energy chirp control.
\newblock {\em Phys. Rev. Lett.}, 117:124801, Sep 2016.

\bibitem{Liu+PRAB-20_2017}
Tao Liu, Tong Zhang, Dong Wang, and Zhirong Huang.
\newblock Compact beam transport system for free-electron lasers driven by a
  laser plasma accelerator.
\newblock {\em Phys. Rev. Accel. Beams}, 20:020701, Feb 2017.

\bibitem{Liu+ProcIPAC_2016}
Tao Liu, Tong Zhang, Dong Wang, Zhirong Huang, and Jiansheng Liu.
\newblock Beam transport line of the lpa-fel facility based on transverse
  gradient undulator.
\newblock In {\em Proceedings of IPAC 2016, Busan, Korea}, 2016.

\bibitem{Afonso+ProcIPAC2011}
V.~Afonso~Rodr\'{i}guez, T.~Baumbach, A.~Bernhard, A.~Keilmann, P.~Peiffer,
  R.~Rossmanith, C.~Widmann, M.~Nicolai, M.~Kaluza, and G.~Fuchert.
\newblock Design optimization for a non-planar undulator for the jeti-laser
  wakefield accelerator in jena.
\newblock In {\em Proceedings of the IPAC2011, San Sebasti\'{a}n, Spain}, 2011.

\bibitem{Afonso+2012}
V.~{Afonso Rodriguez}, A.~Bernhard, A.~Keilmann, P.~Peiffer, R.~Rossmanith,
  C.~Widmann, T.~Baumbach, M.~Nicolai, and M.~C. Kaluza.
\newblock {Development of a Non-Planar Superconducting Undulator for the
  JETI-Laser-Wakefield Accelerator}.
\newblock {\em IEEE Transactions on Applied Superconductivity}, 23(3 Part
  2):4101505, 2013.

\bibitem{Afonso2015}
Ver\'{o}nica {Afonso Rodr\'{i}guez}.
\newblock {\em {Electromagnetic Design, Implementation and Test of a
  Superconducting Undulator with a Transverse Gradient Field Amplitude}}.
\newblock PhD thesis, Karlsruhe Institute of Technology, 2015.

\bibitem{Widmann+2014}
C.~Widmann, V.~{Afonso Rodriguez}, A.~Bernhard, N.~Braun, A.-S. Mueller,
  A.~Papash, R.~Rossmanith, W.~Werner, M.~C. Kaluza, M.~Reuter, M.~Nicolai, and
  A.~S\"{a}vert.
\newblock {Beam transport system from a Laser Wakefield Accelerator to a
  Transverse Gradient Undulator}.
\newblock In {\em {Proceedings of the International Particle Accelerator
  Conference IPAC2014}}, page THOBA03, 2014.

\bibitem{Widmann2016}
Christina Widmann.
\newblock {\em Simulation and first experimental tests of an electron beam
  transport system for a laser wakefield accelerator}.
\newblock PhD thesis, Karlsruhe Institute of Technology, 2016.

\bibitem{Baxevanis+2014}
Panagiotis Baxevanis, Yuantao Ding, Zhirong Huang, and Ronald Ruth.
\newblock {3D theroy of a high-gain free-electron laser based on a transverse
  gradient undulator}.
\newblock {\em Phys. Rev. ST Accel. Beams}, 17:020701, 2014.

\bibitem{Bernhard+PRAB-19_2016}
Axel Bernhard, Nils Braun, Ver\'onica~Afonso Rodr\'{\i}guez, Peter Peiffer,
  Robert Rossmanith, Christina Widmann, and Michael Scheer.
\newblock Radiation emitted by transverse-gradient undulators.
\newblock {\em Phys. Rev. Accel. Beams}, 19:090704, Sep 2016.

\bibitem{Scheer2012}
Michael Scheer.
\newblock {WAVE - A Computer Code for the Tracking of Electrons through
  Magnetic Fields and the Calculation of Spontaneous Synchrotron Radiation}.
\newblock In {\em {Proceedings of ICAP 2012}}, page TUACC2, 2012.

\bibitem{Bernhard+ProcIPAC_2015a}
A.~Bernhard, V.~{Afonso Rodriguez}, J.~Senger, W.~Werner, C.~Widmann, and A.-S.
  M\"{u}ller.
\newblock {Compact In-Vacuum Quadrupoles for a Beam Transport System at a Laser
  Wakefield Accelerator}.
\newblock In {\em {Proceedings of the International Particle Accelerator
  Conference IPAC2015}}, page WEPMA038, 2015.

\bibitem{Hillenbrand+ProcIPAC_2015}
S.~Hillenbrand, A.~Bernhard, A.-S. Mueller, M.~J. Nasse, R.~Rossmanith,
  R.~Ruprecht, M.~Sauter, M.~Schuh, S.~Schulz, M.~Weber, P.~Wesolowski, and
  C.~Widmann.
\newblock {Magnet studies for the FLUTE accelerator at KIT}.
\newblock In {\em {Proceedings of the International Particle Accelerator
  Conference IPAC2015}}, page WEPMA040, 2015.

\bibitem{Will+ProcIPAC_2017}
A.~Will, A.~Bernhard, M.~Kaluza, A.-S. Müller, and C.~Widmann.
\newblock {D}etailed {A}nalysis of a {L}inear {B}eam {T}ransport {L}ine from a
  {L}aser {W}akefield {A}ccelerator to a {T}ransverse{-G}radient {U}ndulator.
\newblock In {\em Proc. of International Particle Accelerator Conference
  (IPAC'17), Copenhagen, Denmark, 2017}, number~8 in International Particle
  Accelerator Conference, pages 1711--1714, Geneva, Switzerland, May 2017.
  JACoW.

\end{thebibliography}
\end{document}